\documentstyle[11pt,newpasp,twoside,epsf]{article}
\def\uchii{\hbox{UCH{\sc ii}}}
\def\hii{\hbox{H{\sc ii}}}

\begin{document}
\title{High-resolution studies of massive star-forming regions 
}
\author{Thomas Henning}
\affil{Astrophysikalisches Institut und Universit\"ats-Sternwarte Jena,
     Schillerg\"a{\ss}chen 3,
     D-07745 Jena, Germany}
\author{Markus Feldt}
\affil{Max-Planck-Institut f\"ur Astronomie, K\"onigsstuhl 17,
D-69117 Heidelberg, Germany}
\author{Bringfried Stecklum}
\affil{Th\"uringer Landessternwarte, Sternwarte 5, D-07778 Tautenburg,
Germany}

\begin{abstract}
We summarize the status of our program of near-infrared adaptive
optics observations of ultracompact \hii{} regions. By means of
selected results we demonstrate the usefulness of this technique
for disentangling the complexity of massive star-forming regions. The
primary scientific aims of our study are the identification of the
ionizing stars, the characterization of the stellar population, the
clarification of the role of dust for the excitation of the \hii{} region
and, last but not least, the search for circumstellar disks surrounding
high-mass stars. We present results obtained by adaptive optics observations
and additional measurements for
G45.45+0.06, G5.89-0.39, and G309.92+0.48 to illustrate our scientific
objectives.

\end{abstract}

\section{Introduction}
With the exception of the Orion Trapezium complex, most Galactic massive star
forming-regions are more distant than 1\,kpc. This large average distance is
one reason to apply high-angular resolution techniques for their study. The
embedded massive stars give rise to ultracompact \hii{} regions (\uchii{}s) -
objects with strong emission lines and scattered radiation (e.g. Churchwell 
1990). The photospheric
flux of stars associated with \uchii{}s is often overwhelmed by the nebular
background, rendering their detection in seeing-limited observations difficult
if not impossible. However, already diffraction-limited imaging at 4-m class telescopes
effectively reduces the background contribution by more than a factor of 10 compared to
conventional observations. This prompted us to apply the technique of adaptive optics (AO)
for the identifaction of the stellar content of \uchii{}s.

Within our program, we used the ESO ADONIS instrument (Beuzit et al. 1994)
at the La Silla 3.6-m
telescope and the MPIA ALFA/Omega-Cass system at the Calar Alto
3.5-m telescope (Feldt et al. 2000). ADONIS was the first astronomical AO system which
became accessible to the community. It utilizes a 256$\times$256 NICMOS array while
the ALFA/Omega-Cass camera is based on a 1k$\times$1k HAWAII detector. 

\section{Target selection}
The current generation of AO systems applies sensing of the wavefront reference
in visible light. The need for a nearby ($\la$ 20\arcsec) natural star is a serious constraint
for the observation of massive young stellar objects which, by their nature, are
deeply embedded. Therefore, we cross-correlated the
USNO-2A catalog (Monet et al. 1996) with compilations of \uchii{}s and methanol maser sources 
(e.g. Wood \& Churchwell 1989, Kurtz et al. 1994, Forster \& Caswell 1998, Norris et al.
1998, Walsh et al. 1998) to identify
potential targets. This led to a sample of objects where a field star is by
chance close to the line of sight. The wavefront reference star has to be bright
enough to yield a meaningful AO correction (Strehl ratio). In a few
cases where no USNO star could be identified, we successfully tried nearby
stars (e.g. G5.89-0.39, G333.6-0.22). Since the magnitudes of the wavefront
reference stars were sometimes close to the limit of the AO system ($V\la$ 13 mag)
and, in addition,
due to the decorrelation of the wavefront with increasing angle from the
science target (on axis), the attainable resolution did not always reach
the diffraction limit but, in any case, was superior to the seeing.
At the typical distance of our targets of 3\,kpc, the spatial resolution at
2.2\,\micron{} corresponds to 600\,AU.
Until now, we acquired data for about 30 objects, some of which, however, are
more evolved \hii{} regions, like Sharpless 106 (Feldt et al. 2001).

\section{Observations}
Almost all sources were observed in the J, H, and K (or K') near-infrared (NIR)
bands. For some targets
we obtained additional imaging using Br$\gamma$, He\,I, and H$_2$ narrow-band
filters. A few objects
were also measured polarimetrically as well as spectroscopically. Since the small
pixel scale of the AO instruments leads to a rather small field of view (FOV,
this holds escpecially for ADONIS), we tried to increase this by applying a
dithering scheme, allowing the determination of the sky contribution for targets
of small spatial extent.

Although we focus on our AO-NIR measurements in the present paper,
supplementary data at NIR, mid-infrared
(MIR) and submm/mm wavelengths have been obtained as well. Especially NIR
imaging polarimetry proved to be a very useful tool for the assessment of the dust
distribution and the source geometry. Furthermore, it allows to locate the
primary illumination source which is
generally the most massive star in the presence of a centro-symmetric polarization
pattern. The 2.2\,\micron{} linear polarization maps shown here are based
on observations using SOFI (Finger et al. 1998) at the ESO-NTT.

\newpage
\section{Results}
In order to illustrate the potential of the AO technique, we present results
for three sources.
\subsection{G45.45+0.06}
This source was among the first targets to be observed with ADONIS 
(Stecklum et al. 1995). Its kinematic distance is 6.6\,kpc and the luminosity amounts to
1.4$\times 10^6\,\rm L_{\sun}$ (Wood \& Churchwell 1989).
The 2.2\,\micron{} image taken with SOFI is shown in Fig.1, together with the
vectors of the linear polarization. At the resolution of 0\farcs7, the source
consists of a compact nebula with a few brightness enhancements. Quite large degrees
of polarization are found in the outer areas, with almost parallel vectors.
This is presumably due to scattering by foreground dust grains,
aligned by the interstellar magnetic field. The foreground polarization will
also disturb the appearance of a centro-symmetric pattern which is presumably
the case for this object. The star at [15\arcsec,10\arcsec]
served as wavefront reference.
The deconvolved ADONIS K´ image is shown on the left of Fig.2 together with
contours of the 6\,cm radio continuum from Wood \& Churchwell (1989). At the
resolution of 0\farcs15, the nebula is resolved into more than a dozen of stars,
several of them arranged in a chain-like fashion along the ionization front.

\begin{figure}
\plotfiddle{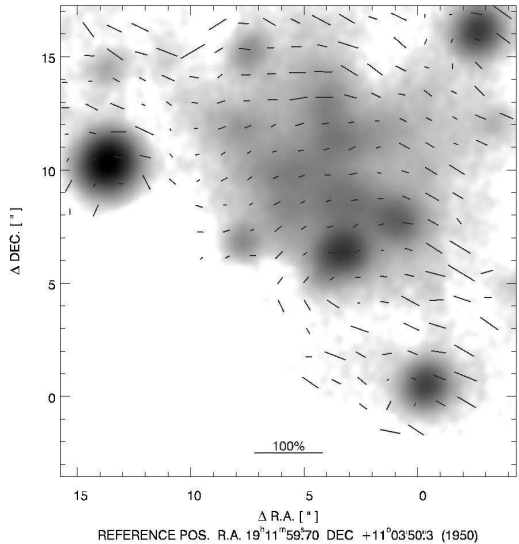}{7.5cm}{0}{40}{40}{-125}{0}
\caption{K' image of G45.45+0.06 with vectors of the linear polarization. The
line at the bottom represents 100\% polarization.}
\end{figure}
\begin{figure}
\plottwo{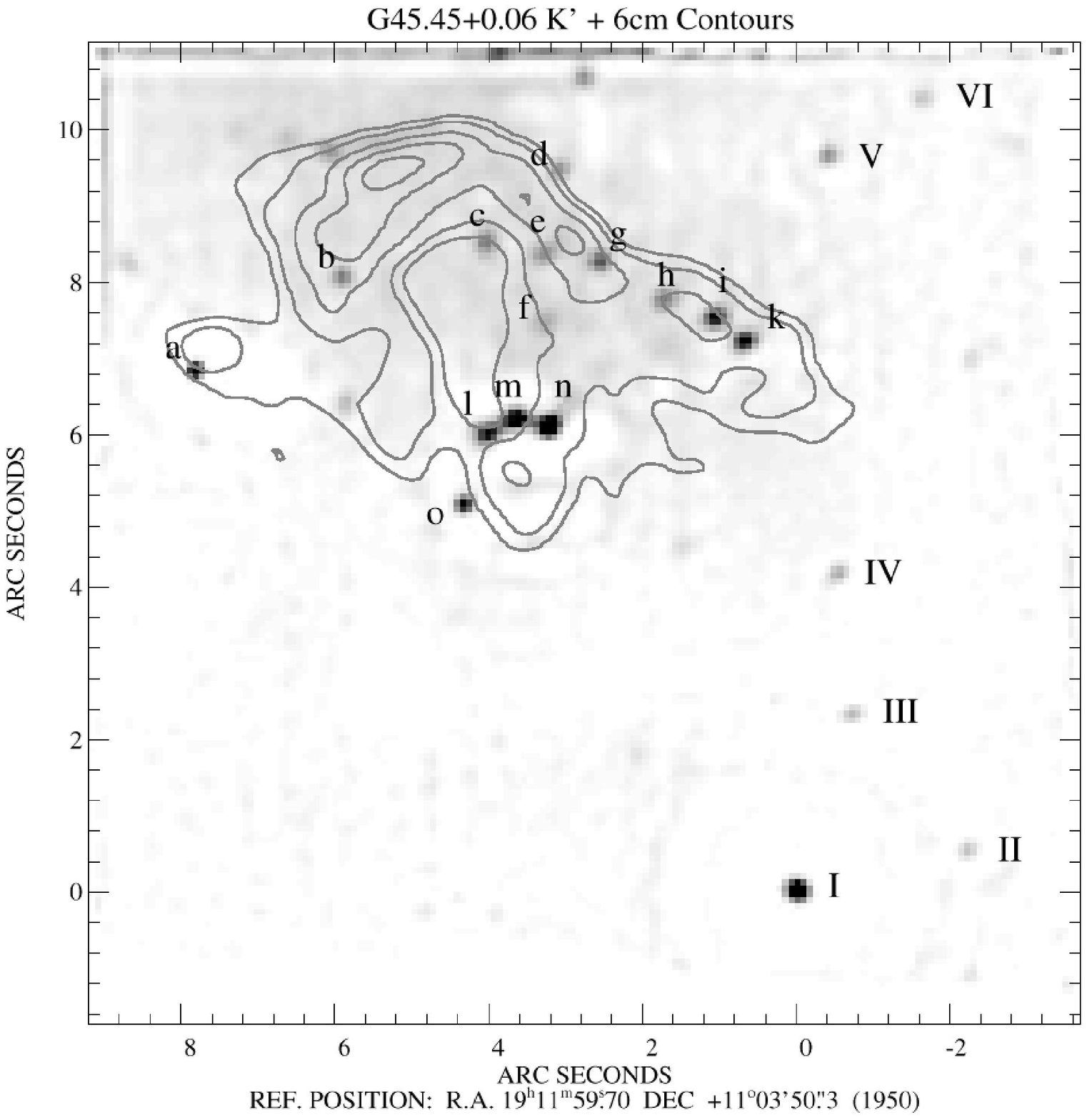}{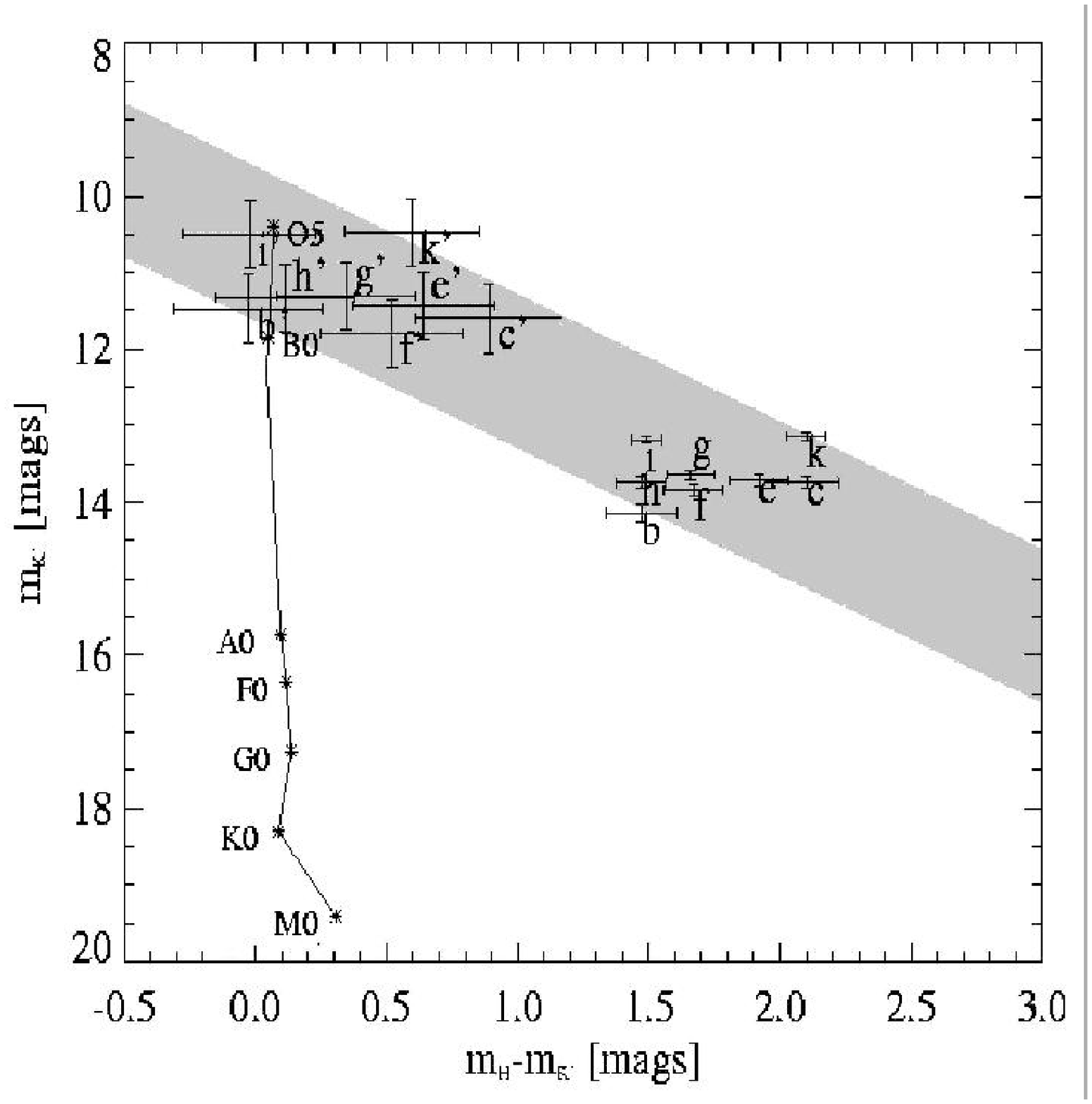}
\caption{Left: Deconvolved ADONIS K' image of G45.45+0.06 with contours of the
6\,cm radio continuum from Wood \&
Churchwell (1989). Right: K--(H--K) color-magnitude diagram with stars shown before
and after reddening. Spectral types along the ZAMS are marked.}
\end{figure}

In order to assess the spectral types of these stars, we estimated the extinction
comparing the observed Br$\gamma$ flux to the expected one based on VLA
radio continuum measurements. The dereddened H and K´ fluxes were used to establish
the K--(H--K) color-magnitude diagram shown in Fig.2 (right). Here, the stars are shown
before and after dereddening (the gray region indicates the reddening path). 
It is obvious that almost all stars fall close to
the upper end of the zero-age-main sequence (ZAMS), proving that they are indeed
massive. The remaining scatter may arise from contributions of circumstellar
extinction or infrared excess due to hot dust. Although we were able
to identify the high-mass stellar population of this \uchii{}, the limited
sensitivity did not allow to study the initial mass function comprehensively,
but demonstrated the presence of a cluster of massive stars.
The detailed analysis of all our data and a thorough discussion of this source
is given by Feldt et al. (1998).

\subsection{G5.89-0.39}
G5.89-0.39 is an archetype \uchii{}, located at the distance of 2.6\,kpc.
Its luminosity of 3.5$\times 10^5\,\rm L_{\sun}$
requires an ionizing star of spectral type O6. The linear polarization map
(Fig.3) shows a very pronounced centro-symmetric pattern, especially
towards the northwest. This implies that NIR radiation can emerge into this
direction rather easily, i.e. at low optical depth. Furthermore, it suggests
the presence of one major illumination source, a single star or a very
compact group of stars. The location of the illuminator can be estimated by
minimizing the sum of the scalar product between the radius and polarization
vectors through varying the position. Using polarization vectors with
polarization degrees in excess of 10\% (single scattering) only, we estimated
the location of the illuminator which is marked by the inclined cross. Notably, there is
no visible stellar source at this position
which immediatly implies a large amount of extinction along this line of sight.  

\begin{figure}
\plotfiddle{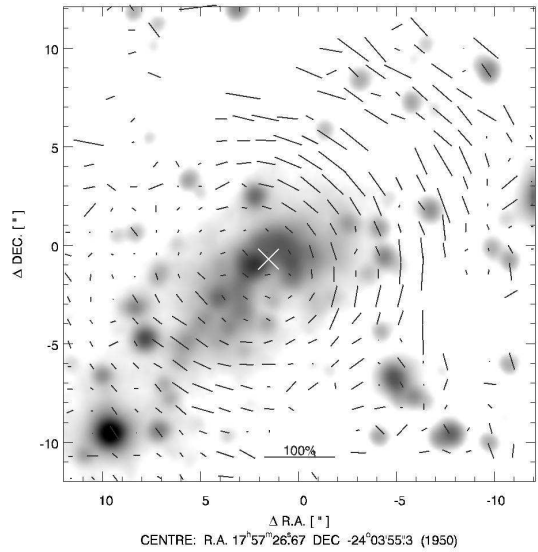}{7.0cm}{0}{40}{40}{-125}{0}
\caption{K' image of G5.89-0.39 with vectors of the linear polarization. The
inclinded cross marks the position of the illuminator.}
\end{figure}
\begin{figure}
\plotfiddle{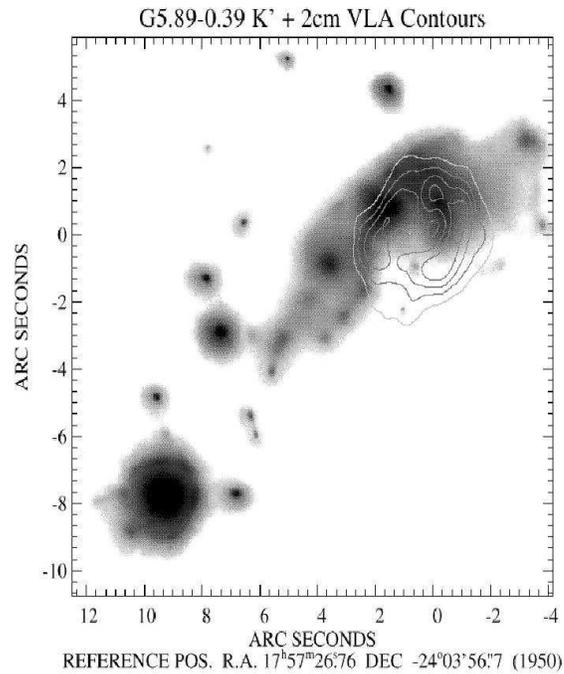}{7.0cm}{0}{39}{39}{-125}{0}
\caption{ADONIS K' image with 2\,cm radio continuum contours from Wood \&
Churchwell (1989)}
\end{figure}

Fig.4 displays the ADONIS K´ image together with the 2\,cm
radio continuum contours from Wood \& Churchwell (1989). While the radio morphology
suggest a shell-like structure, the NIR appearance is asymmetric, with emission
from the northern part only (cf. Harvey et al. 1994). We found that the IR appearance does not
change from the K´ to the Q band, i.e. over an order of magnitude in wavelength!
The lack of any IR recombination line emission from the southern part of the
radio continuum region also implies a large extinction. Indeed, our 1.3\,mm 
dust continuum map obtained with the SEST (and corrected for free-free emission)
shows a dense core at this very location. Thus we conclude that the IR appearance
of G5.89-0.39 is strongly affected by large column densities of foreground dust.
This has been put forward by the analysis of our multi-wavelength data set
(Feldt et al. 1999). 
Our recent ALPHA images suggest that the dust may be contained in a pillar-like
structure, similar to those of M16, and that our line of sight is only slightly
inclined with respect to the symmetry axis of the pillar. Thus, the direct view
on the exciting star is precluded while NIR radiation can emerge and be seen
in the northwest area which is not shadowed by the pillar. In this simplistic
model, the radio continuum would arise from the ionized skin of the pillar and
its temporal variation may reflect the change of its size due to the evaporation.
However, details of this model have to be worked out yet.

\subsection{G309.92+0.48}
This \uchii{} belongs to the group of objects which is associated with methanol
masers, thought to indicate the earliest stages of massive star formation.
Moreover, the chainlike alignment of the maser spots as well as the systematic
trend in radial velocity along this line led Norris et al. (1998) to suggest
that the masers reside in a circumstellar disk surrounding a high-mass star. 
The kinematic distance of the object is 6.3\,kpc and its luminosity amounts to 
4.3$\times 10^5\,\rm L_{\sun}$. Fig.5 shows
the 2.2\,\micron{} SOFI image with superimposed vectors of the linear polarization
as well as contours of the HeI emission based on our ADONIS narrow-band imaging.
For this source, two centro-symmetric patterns were found, one of them centred
on the \uchii{} while the second one is caused by a less embedded star to the
northwest. The location of the illuminator of the first pattern is marked by
the slanted cross while the plus sign marks the reference position (center
of the maser spots). Interestingly, the illuminating source does not
correspond to the brightest 2.2\,\micron{} object but to a star slightly offset
to the southeast with respect to the maser location which is associated with
HeI emission. This proves our conjecture that the most luminous, i.e. hottest
star, dominates the radiation field. Obviously, more photons emitted by this
star are scattered compared to any other source in the region. We emphasize
that immediatly
to the northwest from the reference position, there is a light area, indicating
strong K´ band extinction presumably due to a dense clump.  Thus, we conclude
that the most massive star is situated very close to a dense clump,
causing the ionization seen in IR recombination lines and in the radio
continuum. The masers
are in between the star and the dense clump, presumably in the photodissociation
region. We found no evidence for a disk around the massive star. We also note
that there is a slight offset between the radio continuum peak and the maser
spots in the Phillips et al. (1998) data which is not consistent with the disk
hypothesis. In the latter case, the radio continuum peak should always be
centered on the maser spots. Furthermore, an edge-on disk as claimed by Norris
et al. (1998) would cause an anisotropic radiation field, resulting in a
bipolar polarization pattern (e.g. Fischer, Henning \& Yorke 1994) which is
not observed.

\begin{figure}
\plotfiddle{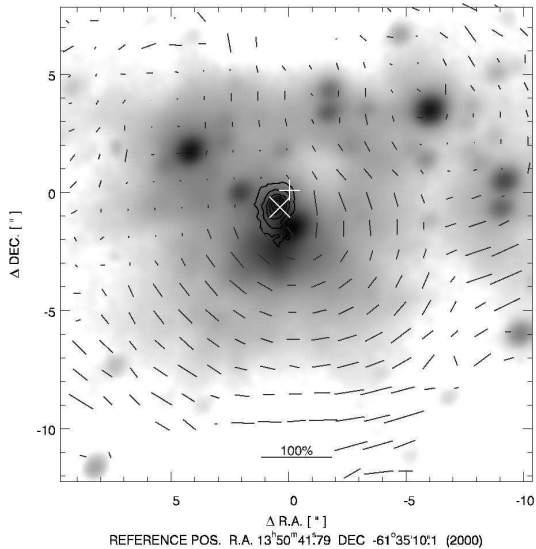}{7.5cm}{0}{40}{40}{-125}{0}
\caption{K' image of G309.92+0.48 with vectors of the linear polarization
and contours of the 2.057\,\micron{} HeI emission. The
inclinded cross marks the position of the illuminator. The plus sign
indicates the methanol maser location (reference position).}
\end{figure}

\section{Conclusions}
For many targets, our AO observations revealed the exciting stars of the \uchii{}.
There is a considerable fraction of objects, however, for which large column
densities preclude the NIR view of the embedded massive stars (e.g. G5.89-0.39,
G9.62+0.19D). The combination of radio continuum maps and NIR images with similar
beam sizes enables the determination of the intrinsic \uchii{} morphology.
Our high-resolution imaging did not yield direct evidence for the presence
of disks around massive stars which were invoked in one model of \uchii{}s
(Hollenbach et al. 1994). The disks might be smaller than our resolution limit
or even be destroyed when the massive star becomes observable in the NIR if
high-mass stars form via disks at all. In almost all cases, however, we were
able to identify dense regions adjacent to the hot stars which may provide
the reservoir to sustain the \uchii{} phase. Furthermore, it became obvious
that line-of-sight effects are important for the IR appearance of \uchii{}s
which, in general, do not obey a simple 1D symmetry.

The next generation of AO instruments which is on the verge, e.g. the ESO system
NAOS/Conica for the VLT, employs infrared wavefront detection.
This will allow an unbiased survey of a large sample of \uchii{}s with
more than a twofold increase in spatial resolution. Instruments like this offer
a great potential for the investigation how massive stars form. But like their
predecessors, they will provide answers as well as questions.

\end{document}